\documentclass[aps,preprint,amsmath,amssymb,superscriptaddress,floatfix]{revtex4-1}
\usepackage{epsfig}
\usepackage{epstopdf}
\usepackage{graphicx}
\usepackage{dcolumn}
\usepackage{color}
\usepackage{bm}
\usepackage{mathrsfs}
\usepackage{bbold}
\usepackage{amssymb}
\usepackage{upgreek}
\usepackage{caption}
\usepackage{subcaption}
\DeclareMathOperator{\Tr}{Tr}

\begin{document}

\title{\Large \bf Quantum thermodynamics of single particle systems}

\author{Md.~Manirul Ali}
\email{maniquantum@gmail.com}
\affiliation{Department of Physics, National Cheng Kung University, Tainan 70101, Taiwan}
\author{Wei-Ming Huang}
\affiliation{Department of Physics, National Cheng Kung University, Tainan 70101, Taiwan}
\author{Wei-Min Zhang}
\email{wzhang@mail.ncku.edu.tw}
\affiliation{Department of Physics, National Cheng Kung University, Tainan 70101, Taiwan}

\begin{abstract}
\textbf{Classical thermodynamics is built with the concept of equilibrium states. However, it is less
clear how equilibrium thermodynamics emerges through the dynamics that follows the principle of
quantum mechanics. In this paper, we develop a theory to study the exact nonequilibrium thermodynamics
of quantum systems that is applicable to arbitrary small systems, even for single particle systems,
in contact with a reservoir. We generalize the concept of temperature into nonequilibrium regime that depends
on the detailed dynamics of quantum states. When we apply the theory to the cavity
system and the two-level atomic system interacting with a heat reservoir, the exact nonequilibrium theory 
unravels unambiguously (1) the emergence of classical thermodynamics from quantum
dynamics in the weak system-reservoir coupling regime, without introducing
equilibrium hypothesis; (2) the breakdown of classical thermodynamics in the strong coupling
regime, which is induced by non-Markovian memory dynamics; and (3) the occurrence
of dynamical quantum phase transition characterized by inflationary dynamics associated 
with a negative nonequilibrium temperature, from which the third law of thermodynamics, allocated in the deep 
quantum realm, is naturally proved. The corresponding  dynamical criticality provides the border separating the classical 
and quantum thermodynamics. The inflationary dynamics may also provide a simple picture for 
the origin of big bang and universe inflation.}
\end{abstract}

\maketitle

\vskip 0.5cm
\noindent
{\large \bf Introduction}
\vskip 0.2cm
\noindent
Recent development in the field of open quantum systems initiates a renewed interest on
the issue of quantum thermodynamics taking place under
nonequilibrium situations \cite{TrotzkyNatPhys12,GringScience12,LangenNatPhys13,Kosloff13,
EisertNatPhys15,Complexity,JMillenNJP16,EspositoNJP17}. Meanwhile, a conceptual
understanding of thermodynamic laws in the quantum domain has also been central to research for
quantum thermodynamics 
\cite{GemmerMahler,Hanggi2011,Jarzynski2011,Scullypnas2011,Sairaprl2012,Esposito2015,Pekola2015,Ronagel2016}.
The questions of how classical thermodynamics emerges from quantum dynamics, how do quantum systems
dynamically equilibrate and thermalize, and whether thermalization is always reachable in quantum systems,  
have attracted much attention in recent years. However, the theory of  quantum thermodynamics that  
has conceptually no ambigurity and can be widely acceptable has yet been reached from various  
approximation methods, due to lack of the exact solution of nonequilibrium dynamics for open quantum systems. 
In this paper, we will answer these questions by rigorously and exactly solving the nonequilibrium dynamics 
based on the exact master equation we developed recently for a large class of open quantum
systems \cite{general,Tu2008,Jin2010,Lei2012,zhang2018}.

Recall that classical thermodynamics is built with the concept of {\it equilibrium states} that are completely
characterized by specifying the relation between the internal energy $E$ and a set of
other extensive parameters: the entropy $S$, the volume $V$, and the particle number $N_i$
of different components $i=1, 2, \cdots$, etc. Explicitly, classical thermodynamics is described
by {\it the thermodynamic fundamental equation}, $E=E(S, V, N_1, N_2, \cdots)$, under the
extremum principle of either maximizing the entropy or minimizing the energy \cite{HBCallen}.
The thermodynamic temperature $T$, the pressure $P$ and the chemical potential $\mu_1, \mu_2, \cdots$ 
are defined by the first derivative of energy with respect to entropy, the volume and the particle number: 
$T=\frac{\partial E}{\partial S}\big)_{V, N_1, N_2, \cdots}$, $P=\frac{\partial E}{\partial V}\big)_{S, N_1, N_2, \cdots}$,
$\mu_1=\frac{\partial E}{\partial N_1}\big)_{S,V, N_2, \cdots}$, $\cdots$ such that $dE= TdS + PdV + \mu_1 dN_1 + \cdots$.
Microscopically, statistical averages of various thermodynamic quantities in equilibrium thermodynamics 
are calculated over a probability distribution function of canonical or grant canonical ensemble in the 
thermodynamic limit (the system with a large number of particles coupled very weakly to a reservoir
at a relatively high termature).

Contrary to the above classical thermodynamics, here we consider the thermodynamics of {\it single quantum 
systems} coupled strongly or weakly to the reservoir at arbitrary temperature, including the zero-temperature
(the deep quantum realm). Then the internal energy $E$ should depend only on one extensive parameter, 
the entropy $S$ of the system, and all other extensive parameters $V, N_1, \cdots $~are absent for such 
simple open quantum systems.  Correspondingly, {\it the fundamental equation of thermodynamics
for such a simple quantum system} takes, in principle, the form
$E = E(S)$.
Conventionally, there is no classical concept of single particle temperature,
whereas the single quantum particle coupled with a reservoir can have ensemble of various microstates validating the
existence of entropy. Thus, the temperature can always be introduced by the definition, $T\equiv \partial E/\partial S$,
from the above fundamental equation.

In this paper, we begin with the exact nonequilibrium quantum dynamics where the system is coupled
to a reservoir. We investigate the emerging
thermodynamics through the real-time dynamics of various properly defined thermodynamic quantities,
i.e., energy, entropy, temperature, and free energy, much before the
system approaches to the steady state. More specifically, we establish the quantum thermodynamics
of the system fully within the framework of quantum mechanics.

\begin{figure}
\includegraphics[width=7.5cm]{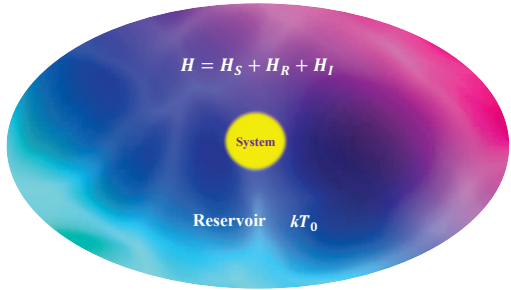}
\caption{\label{SchmDiag} {\bf Schematic illustration of an open quantum system.} An open quantum system
described by the Hamiltonian $H_S$ is in contact with a heat reservoir with Hamiltonian $H_R$, and $H_{I}$
describes the interaction between the system and the reservoir. The reservoir is taken initially
in a thermal equilibrium state with an initial temperature $T_0$.}
\end{figure}
\vskip 0.5cm
\noindent
{\large \bf Results}
\vskip 0.2cm
\noindent{\bf 1.~Formulation of quantum thermodynamics.} Consider a standard open quantum
system coupled to an arbitrary reservoir that is described by the Hamiltonian,
\begin{align}
H = H_S + H_R + H_I, 
\end{align}
which consists of the Hamiltonians of the system and the reservoir, denoted
respectively by $H_S$ and $H_R$, and the interaction between them, $H_I$.
The general dynamics of the open quantum system can be determined by the exact master equation of
the reduced density matrix $\rho_S(t)$, which is defined by tracing over all the reservoir degrees
of freedom from the total density matrix $\rho_{tot}(t)$ of the system plus reservoir: $\rho_S(t)=\Tr_R[\rho_{tot}(t)]$.
The total density matrix $\rho_{tot}(t)$ obeys the von Neumann equation \cite{Neumann55}: $i\hbar \dot{\rho}_{tot}(t)=[H, \rho_{tot}(t)]$.
Thus, the formal solution of the von Neumann equation gives
\begin{align}
\rho_S(t)= \Tr_R[U(t,t_0)\rho_{tot}(t_0)U^\dag(t,t_0)],
\end{align}
where the time-evolution operator $U(t,t_0)=\exp\{\frac{1}{i\hbar}H(t-t_0)\}$, and $\rho_{tot}(t_0)$ is the
initial state of the total system (the system plus the reservoir).
The reduced density matrix $\rho_S(t)$ provides all the dynamics of the system
under the time evolution and obey the exact master equation for a large class of open
quantum systems given in Refs.~\cite{general,Tu2008,Jin2010,Lei2012,zhang2018}.  
From this exact formalism, we can use the von Neumann definition of entropy
to define the nonequilibrium entropy of the system, which provides in general the lower bound of the 
entropy of the system and it is also the entanglement entropy between the system and the reservoir,
\begin{eqnarray}
S(t) = - k \Tr_{S} \left[ \rho_S(t) \log \rho_S(t) \right].
\label{vontropy}
\end{eqnarray}
Here we have introduced the Boltzmann constant $k$ in the above definition, in order to
make a clear connection to the thermal entropy in the later discussion. Whether or not this 
nonequilibrium entropy can be properly defined as the quantum thermodynamic entropy 
lies on the fact whether it can correctly reproduce the equilibrium entropy in classical 
thermodynamics when the system reaches the equilibrium state.
On the other hand, the nonequilibrium internal energy of the system at time $t$ is quantum mechanically given by
\begin{eqnarray}
E(t) = \Tr_{S} \left[  \rho_S(t) H_S \right],
\label{avengy}
\end{eqnarray}
which is also fully determined by $\rho_S(t)$. Thus, equations (\ref{vontropy}-\ref{avengy})
establishes {\it the fundamental relation of nonequilibrium quantum thermodynamics}
between the energy and entropy through the quantum density matrix $\rho_S(t)$
of open quantum systems under a rigorous nonequilibrium description.

Based on the above fundamental relations for nonequilibrium quantum thermodynamics, we generalize
the concept of temperature far-beyond equilibrium. We introduce a temperature-like
quantity ${\cal T}(t)$ which is defined as the derivative of the nonequilibrium internal energy
with respect to the nonequilibrium entropy of the system as follows:
\begin{eqnarray}
{\cal T}(t) \equiv \frac{\partial E(t)}{\partial S(t)} 
\label{kT2}
\end{eqnarray}
We call this quantity as the nonequilibrium temperature (or dynamical temperature) of the system because it evolves through various
quantum states much before it approaches to the steady state or thermal equilibrium. Conditionally,
the nonequilibrium temperature defined by Eq.~(\ref{kT2}) is physically meaningful if it is coincident to the
thermal equilibrium temperature in the steady-state limit when the system approaches the thermal equilibrium state
in the weak-coupling regime, as we will show later in the paper.

Consequently, using Legendre transformation of $E(t)$ with respect to $S(t)$ that replaces the entropy
by the  nonequilibrium temperature as the independent variable, we obtained the nonequilibrium Helmholtz  free energy,
\begin{eqnarray}
F(t) = E(t) - {\cal T}(t) S(t),
\label{fengy}
\end{eqnarray}
which is a nonequilibrium quantum thermodynamic potential resulting from the time evolution of
the system. Based on such nonequilibrium quantum thermodynamic potential $F(t)$, one
can define the nonequilibrium quantum thermodynamic work, which is the time-dependent
dissipative work done by the system,
as the decrease of the nonequilibrium Helmholtz free energy of the system: $dW(t) = - dF(t)$.
Thus, the time-dependent (nonequilibrium) heat transferred into the system
$dQ(t)$ can be obtained as a consequence of the energy conservation,
\begin{eqnarray}
dQ(t) = dE(t) - dF(t),
\label{work}
\end{eqnarray}
where $dE(t)$ is the change of the internal energy in the nonequilibrium process.

With this new formulation of nonequilibrium quantum thermodynamics,
we show next (i) how the classical thermodynamics emerges from quantum dynamics of single particle
systems in the weak system-reservoir coupling regime, without introducing any hypothesis on
equilibrium state; (ii) how the classical thermodynamics breaks down in the strong coupling regime
due to non-Markovian memory dynamics; and (iii) a dynamical quantum phase transition associated
with the occurrence of negative temperature when the reservoir is initially in vacuum state or its
{\it initial} thermal energy $kT_0$ is less than the system initial energy $E(t_0)$. The dynamical criticality 
in this nonequilibrium quantum phase transition provides a inflationary dynamics that may
reveal a simple physical picture to the origin of universe inflation \cite{Guth1981}.
We also show how the third law of thermodynamics, as a true quantum effect, is obtained in this theory
of quantum thermodynamics.

\vskip 0.2cm
\noindent
{\bf 2.~Application to a cavity system.}
We first consider the system as  a single-mode photonic cavity
with $H_S=\hbar \omega_s a^{\dagger} a$, in contact with a heat bath
(it is also applicable to fermionic system in a similar way \cite{Tu2008,Jin2010}). Here $\omega_s$ is the frequency of the photon field
in the cavity.  The bath is modeled as a collection of infinite photonic modes with a continuous frequency spectrum,
and the system-bath interaction is described
by the Fano-Anderson model that has wide applications in atomic physics and
condensed matter physics \cite{FanoAnd1,FanoAnd2,FanoAnd3}.  Thus, the total Hamiltonian of the system plus the bath is given by
\begin{align}
H= \hbar \omega_s a^\dag a + \sum_k \hbar \omega_k b^\dag_k b_k + \sum_k \hbar (V_k a^\dag b_k + V^*_k b^\dag_k a) .
\end{align}
The reduced density matrix of the
system obey the following exact master equation  \cite{general,Lei2012,Xiong2010}:
\begin{align}
\frac{d}{d t} \rho_S(t)  =  -i  \omega_s^{\prime}(t,t_0)\big[a^\dag a ,  \rho(t)\big]  + \gamma(t,t_0)
\big[2a \rho_S(t) a^{\dag} -\! a^{\dag} a \rho_S(t) \! -\! \rho_S(t)
a^{\dag}a\big]  ~ \notag \\  +\widetilde{\gamma}(t,t_0) \big [a^{\dag}\rho_S(t) a
+ a\rho_S(t)  a^{\dag} - a^{\dag}a\rho_S(t) \! -\!  \rho_S(t) a a^{\dag}\big],
\label{Exact-ME}
\end{align}
in which the renormalized system frequency, the dissipation (relaxation) and fluctuation (noise) coefficients
are determined exactly by  $\omega_s^{\prime}(t,t_0) = -{\rm Im}[\dot{u}(t,t_0)/u(t,t_0)]$,
$\gamma(t,t_0) = -{\rm Re}[\dot{u}(t,t_0)/u(t,t_0)]$
and $\widetilde{\gamma}(t,t_0)=\dot{v}(t,t)-2v(t,t){\rm Re}[\dot{u}(t,t_0)/u(t,t_0)]$. These exact
coefficients are determined by the two basic Schwinger-Keldysh nonequilibrium Green' function
$u(t,t_0)=\langle [a(t),a^{\dagger}(t_0)] \rangle$, and $v(t,t^{\prime}) \sim \langle a^{\dagger}(t^{\prime})
a(t) \rangle$ (apart by an initial-state dependent term) that satisfy the following integro-differential
equations \cite{general,Lei2012}:
\begin{subequations}
\label{uvt}
\begin{align}
&\frac{d}{dt}u(t,t_0) + i \omega_s u(t,t_0) + \int_{t_0}^t d\tau g(t,\tau) u(\tau,t_0) = 0. \label{ide} \\
&v(t,t)=\int_{t_0}^t d\tau_1 \int_{t_0}^t d\tau_2 ~ u(t,\tau_1) {\widetilde g}(\tau_1,\tau_2)u^{\ast}(t,\tau_2), \label{vtb}
\end{align}
\end{subequations}
The integral kernels $g(t,\tau) = \int_0^{\infty} d\omega  J(\omega) e^{-i\omega(t-\tau)}$ and
${\widetilde g}(\tau_1,\tau_2)\!=\!\int_0^{\infty} d\omega  J(\omega) {\overline n}(\omega,T_0) e^{-i\omega(\tau_1-\tau_2)}$
are determined by the spectral density $J(\omega)$ of the reservoir, and $\overline{n}(\omega,T_0)=\frac{1}{e^{\hbar
\omega / k T_0}-1}$ is the initial particle number distribution of the reservoir. The Green's function
$v(t,t)$ which characterizes the nonequilibrium thermal fluctuations gives the nonequilibrium
fluctuation-dissipation theorem \cite{general}.
The exact analytic solution of the nonequilibrium propagating Green function $u(t,t_0)$
has been recently given  \cite{general}
as $u(t,t_0)=\int_{-\infty}^{\infty} d\omega {\mathcal D}(\omega) \exp \{-i\omega(t-t_0) \}$
with ${\mathcal D} (\omega) = {\mathcal D}_l(\omega) + {\mathcal D}_c(\omega)$, where
${\mathcal D}_l(\omega)$ is the contribution of a dissipationless localized bound state in the Fano model,
and ${\mathcal D}_c(\omega)$ is the continuous part of the spectra (corresponding to the level
broadening induced by the reservoir)
that characterizes the relaxation of the system. Without loss of generality, we consider an Ohmic spectral density
$J(\omega) = \eta \omega \exp\left(-\omega/\omega_c \right)$, where $\eta$ is the coupling strength between the
system and the thermal reservoir, and $\omega_c$ is the frequency cutoff of the reservoir spectra \cite{LeggettRMP}.
In this case, the system has a localized bound state when the coupling strength $\eta$ exceeds the critical value
$\eta_c=\omega_s/\omega_c$. With the above specification, all quantum thermodynamical quantities can
be calculated exactly.

Explicitly, let the system be prepared in an energy eigenstate with the initial density matrix $\rho(t_0)=
|E_0 \rangle\langle E_0|$, where $E_0 = {n_0} \hbar \omega_s$ and ${n_0}$ is an integer.
The reservoir is initially in thermal equilibrium state with an initial temperature $T_0$. {\it After the initial time $t_0$,
both the system and the reservoir evolve into nonequilibrium states}. The system state at
arbitrary later time $t$ can be obtained rigorously by solving the exact master equation of Eq.~(\ref{Exact-ME}).
The result is \cite{BreakBE}:
\begin{subequations}
\label{brho}
\begin{align}
&\rho_S(t) =  \sum_{n=0}^{\infty} W^{n_0}_n(t) |E_n \rangle \langle E_n|,  \\
&W_{n} ^{n _{0}} (t) = \frac{[ v (t,t)] ^{n}}{[ 1 +
v (t,t) ] ^{n + 1}} \left[ 1 - A \left( t,t_0 \right) \right] ^{n _{0}}\notag \\
&~~~~~~~~~~~~~~~ \times \!\!\!\!\!\!\sum_{k=0}^{\rm min\{n_0,n\}}\!\!\!\Bigg(\!\!\begin{array}{c} n_0 \\ k \end{array}\!\!\Bigg)
\Bigg(\!\!\begin{array}{c} n \\ k \end{array}\!\!\Bigg)\Bigg[\frac{1}{v(t,t)}\frac{A(t,t_0)}{1-A(t,t_0)}
\Bigg]^k , \label{rho_Fock}
\end{align}
\end{subequations}
where $A \left( t,t_0 \right) = \frac{\left| u \left( t , t _{0} \right) \right| ^{2}}{1 + v \left( t , t \right)}$.
From these results, all the nonequilibrium thermodynamic quantities, namely the time-dependent
energy, entropy, temperature and free energy, etc., can be calculated straightforward.
Explicitly, the function $W^{n_0}_n(t)$ is the probability of finding the quantum system with energy
$E_n=n\hbar \omega_s$ at time $t$. The  internal energy Eq.~(\ref{avengy}) of the system at time
$t$ is given by $E(t) = \sum_n W^{n_0}_n(t) E_n$.
The von Neumann entropy of the density operator $\rho(t)$, i.e., Eq.~(\ref{vontropy}), is simply reduced to
$S(t) = - k \sum_{n=0}^{\infty} W^{n_0}_n(t) \log W^{n_0}_n(t)$.
These results establish explicitly the fundamental relation between the nonequilibrium
energy and entropy of the system in the last section, from which one can determine the nonequilibrium temperature
${\cal T}(t)$ defined by Eq.~(\ref{kT2}), and all other related nonequilibrium quantum thermodynamical quantities.
The results are given as follows.

\vskip 0.2cm
\noindent {\bf (i).\it~Emergence of thermodynamics and thermo-statistics from quantum dynamics
in the classical condition $kT_0 > E_0$}.
It is indeed straightforward to see how
the system is dynamically thermalized in the weak system-reservoir coupling regime $(\eta \ll \eta_c)$.
Initially the system is at zero temperature (because it is in the pure state $|E_0\rangle$) with zero entropy,
and the reservoir is in thermal equilibrium at a relatively high temperature $T_0$ with $kT_0 > E_0=n_0\hbar\omega_s$.
Then the system and the reservoir undergo a nonequilibrium evolution due to their contact with each other. The nonequilibrium
dynamics is fully determined by the nonequilibrium Green functions $u(t,t_0)$ and $v(t,t)$ of
Eq.~(\ref{uvt}). One can indeed show explicitly that in the long time steady-state limit $t\rightarrow t_s$,
we have $u(t_s, t_0) \rightarrow 0$, and the average photon number in the cavity obey the Bose-Einstein
statistical distribution,
\begin{align}
\langle a^\dag(t_s)a(t_s) = v(t_s, t_s) \rightarrow  \overline{n}(\omega_s,T_0)
=\frac{1}{e^{\hbar \omega_s/kT_0}-1} .
\end{align}
The probability $W^{n_0}_n(t \rightarrow t_s)$ of occupying the $n$-th
energy state is then solely determined by  the Bose-Einstein statistical distribution,
$\overline{n}(\omega_s,T_0)$, and the system state ultimately approaches unambiguously
to the thermal equilibrium with the reservoir \cite{BreakBE,Complexity}
\begin{eqnarray}
\label{Gibbs}
\rho(t\rightarrow t_s) = \sum_{n=0}^{\infty} \frac{[\overline{n}(\omega_s,T_0)]^n}{[1+\overline{n}(\omega_s,T_0)]^{n+1}}
|E_n\rangle \langle E_n|=\frac{1}{Z}e^{-\frac{H_S}{k T_0}}
\end{eqnarray}
where $Z=\Tr_S[\exp(-H_S/{kT_0})]$, and $T_0$ is the initial temperature of the reservoir.  This provides a rigorous
proof of the dynamical thermalization within quantum mechanical framework.

\begin{figure}[t]
\includegraphics[width=11.0cm]{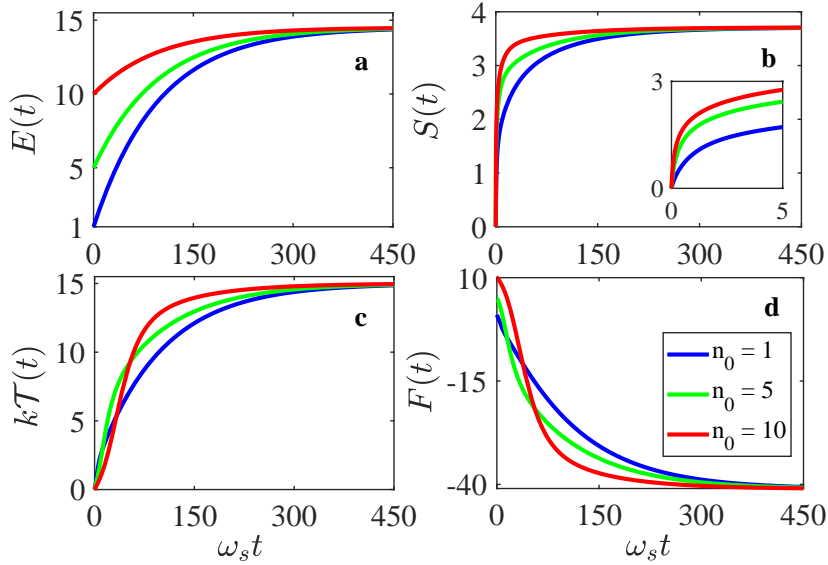}
\caption{\label{Weak} {\bf Nonequilibrium quantum thermodynamics of a photonic cavity}.
The nonequilibrium dynamics of the quantum system in terms of various thermodynamic quantities
in the weak system-reservoir coupling regime with different initial states. Different curves represent different
initial states: $|E_0\rangle=|E_{n_0}\rangle$ with $n_0=1, 5$ and $10$. The initial temperature of the thermal reservoir
is taken at $kT_0=15 \hbar \omega_s$, the cut-off frequency $\omega_c=5\omega_s$, and the system-reservoir
coupling constant $\eta=0.01 \eta_c$.}
\end{figure}

One can then calculate the time-dependent internal energy and the entropy.
In Fig.~\ref{Weak}, we show the dynamics of nonequilibrium internal energy $E(t)$,
entropy $S(t)$,  temperature ${\cal T}(t)$ and  free energy $F(t)$ in the weak
system-reservoir coupling regime for several different initial states of the system.
As we see, both the energy and entropy of the system increases with time due to
its contact with the thermal reservoir through a nonequilibrium process (see
Fig.~\ref{Weak}a and \ref{Weak}b).  The nonequilibrium temperature
${\cal T}(t)$ of the system increases gradually (Fig.~\ref{Weak}c) and
finally approaches to the temperature that is identical to the temperature $T_0$ of
the reservoir, namely the system reaches the equilibrium with the reservoir.

\begin{figure}[t]
\includegraphics[width=11.0cm]{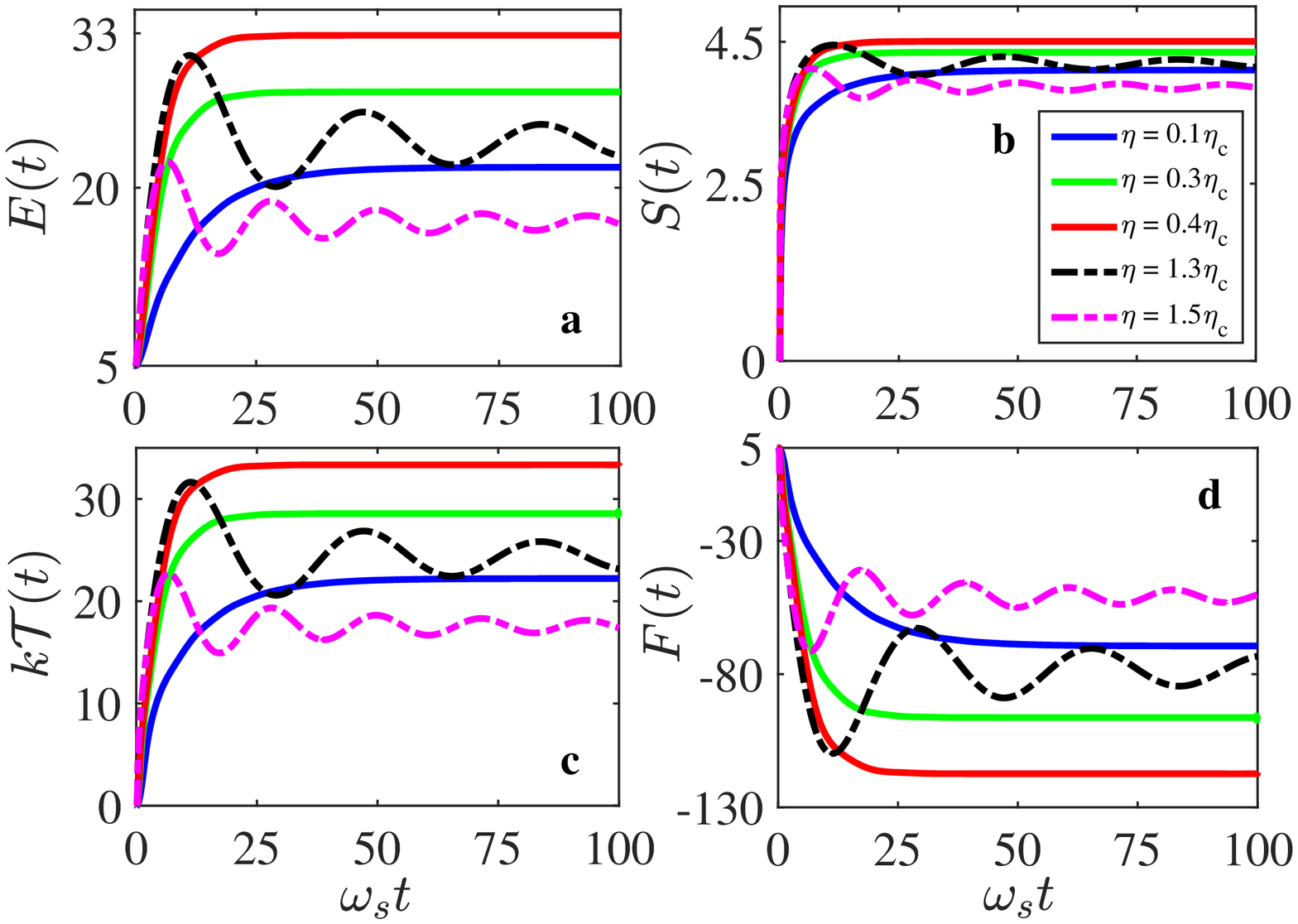}
\caption{\label{Strong} {\bf Nonequilibrium quantum thermodynamics of a photonic cavity in the strong-coupling regime}.
The nonequilibrium dynamics of the quantum system in terms of
various thermodynamic quantities in the strong system-reservoir coupling regime. Different curves represent
different values of the coupling strengths $\eta=0.1\eta_c$ (blue), $\eta=0.3\eta_c$ (green),
$\eta=0.4\eta_c$ (red), $\eta=1.3\eta_c$ (black), $\eta=1.5\eta_c$ (magenta). The other parameters
are $\omega_c = 5 \omega_s$, $kT_0=20 \hbar \omega_s$, and the system is initially in
the state $|E_0\rangle=|E_5\rangle$.}
\end{figure}

The results in Fig.~\ref{Weak}
show precisely how thermalization is dynamically approached in the weak system-reservoir
coupling for a single quantum system, {\it independent of its initial states}.
As it is expected, the entropy always increases in nonequilibrium processes [see  Fig.~\ref{Weak}b],
demonstrating the second law of thermodynamics. The second law places constraints on state transformations
\cite{OppenheimPNAS}, according to which the system evolving nonequilibriumly from the initial state $\rho(t_0)$
makes the free energy $F(t)$ goes down (i.e., delivering work by the system), as shown in Fig.~\ref{Weak}d.
Meantime, the increasing internal energy with the decreasing free energy (work done by the system) always makes
the heat transfer {\it into} the system (nonequilibrium extension of the first law of thermodynamics),
resulting in the increase of the nonequilibrium temperature ${\cal T}(t)$, as shown in Fig.~\ref{Weak}c.
In the stead-state limit, the free energy approaches to a minimum value with maximum entropy in the equilibrium state.
{\it This solution in the weak system-reservoir coupling regime produces
precisely the whole theory of classical thermodynamics without introducing any hypothesis
on equilibrium state of the system from the beginning.}  It is also very different from the previous 
investigations with Born-Markovian approximation, where the reservoir is required to remain 
in the equilibrium state in all the time. Indeed, reproducing the classical thermodynamics in the classical 
regime from the exact nonequilibrium quantum thermodynamics with arbitrary 
initial states of the system is the necessary condition for unambiguously formulating quantum 
thermodynamics.

\vskip 0.2cm
\noindent {\bf (ii).\it~Breakdown of thermodynamics and thermo-statistics in strong system-reservoir coupling regime}.
In fact, the above thermalization crucially relies on the assumption of the very weak interaction
between the system and the reservoir.
In the literature, there is a specific thermalization condition on the system-reservoir coupling \cite{EisertPRL12}: 
the strength of the coupling $\eta$ induces the effect that has to be negligible in
comparison to initial thermal energy $kT_0$. We find that thermalization occurs when $ \eta \ll \eta_c$, where $\eta_c$
is the critical value of the system-reservoir coupling for the occurrence of  localized bound state \cite{general}. In particular,
our numerical calculation shows that the ideal thermalization condition occurs when $\eta \lesssim 0.01 \eta_c$.
On the other hand, the role of strong coupling on equilibrium quantum thermodynamics was only discussed
in the context of damped harmonic oscillators for quantum Brownian motion
\cite{StrongOscillator,StrongQBM}. Some efforts have been made to formulate
quantum thermodynamics for a strongly coupled system
\cite{StrongNoneq1,StrongNoneq2,StrongNoneq3} under nonequilibrium
conditions subject to an external time-dependent perturbation. However, the real-time
dynamics of thermodynamic quantities were not explored from a full quantum dynamics
description in the strong system-reservoir coupling regime.

We find that the real-time quantum thermodynamics of the strong system-reservoir coupling is indeed quite distinct.
As it is shown in Fig.~\ref{Strong}, when the system-reservoir coupling strengths are not so weak
(in the intermediate range: $0.01 \eta_c < \eta \leqslant \eta_c$), the system will quickly approaches to different
thermal equilibrium states (called thermal-like states in Ref.~\cite{Complexity}) for different coupling strengths
with different steady-state temperatures that are larger than the initial reservoir
temperature $T_0$ (see the blue, green and
red curves in Fig.~\ref{Strong}c that correspond to different system-reservoir coupling
strengths). For this intermediate coupling strength region, different equilibrium states
are reached by different steady-state minima of the free energy $F(t)$. These results already
show some deviation away from the
classical thermodynamics prediction, namely after a long enough time, the system must
equilibrate with the reservoir that is characterized by the reservoir temperature $T_0$.
In fact, due to the relatively stronger couplings between the system and the reservoir, both
the system and the reservoir undergo a nonequilibrium evolution, and then approach to
a new common equilibrium state
characterized by the new equilibrium temperature ${\cal T}(t \rightarrow t_s)$.
In Fig.~\ref{Strong}c, it shows that $t_s \simeq 25/\omega_s$, and ${\cal T}(t_s)$
changes as $\eta$ varies. 
But the principle of thermodynamics, namely the existence of equilibrium state and the
associated thermodynamic laws in the steady state, is still valid.

Furthermore, when the system-reservoir coupling strength gets very  strong (goes beyond the critical value $\eta > \eta_c$),
the system indeed does not approach to thermal equilibrium \cite{Complexity}. In this strong coupling
regime, the average energy $E(t)$ initially increases, and then it starts decreasing with time, indicating a backflow of
energy (see Fig.~\ref{Strong}a) from the system into the reservoir, and then it oscillates.
A similar dynamical behavior of entropy $S(t)$ is seen (Fig.~\ref{Strong}b) under
this strong coupling, namely, although the entropy increases in the short time,  it is decreased and
oscillates in later time. Thus, strong system-reservoir couplings cause significant departure from the classical thermalization.
We also {\it do not} see (Fig.~\ref{Strong}c) a monotonous increase of dynamical temperature
${\cal T}(t)$ that can eventually approach to a steady-state value.
In the strong system-reservoir coupling regime, the system indeed cannot reach to a thermal equilibrium, insteadly it approaches
to a more complex nonequilibrium (oscillatory) steady state \cite{Complexity}.
Under this situation, the nonequilibrium free energy $F(t)$ also shows
nontrivial oscillations with local minima, instead of having a global minimum, as shown in Fig.~\ref{Strong}d.

The physical origin for the breakdown of classical thermodynamics in the strong system-reservoir coupling regime
comes from the important effect of nonequilibrium dynamics, i.e. the non-Markovian back-action memory effect, 
as we have discovered earlier \cite{general,Complexity}. In this situation, the system can memorize forever some of its initial state information, i.e.
initial state information cannot be completely wiped out. Therefore, the basic hypothesis of statistical mechanics,
{\it namely after a long enough time, the system will evolve into an equilibrium state with the reservoir that is
independent of its initial state},  breaks down and the system cannot be thermalized. The oscillation behavior in
Fig.~\ref{Strong} is a manifestation of non-Markovian memory effect originating from the localized
bound state of the system, a general non-Markovian dynamic feature in open quantum systems \cite{general}.
The localized bound states of the system produces dissipationless dynamics that do not allow the system to
be thermalized with its environment, which was noticed earlier by Anderson \cite{FanoAnd1,FanoAnd3}
and is justified recently by one of us \cite{Complexity}. The localized bound state can
appear when the system-environment coupling strength exceeds some critical value $\eta_c$ that is determined
by the structure of spectral density $J(\omega)$. For the Ohmic spectral density, $\eta_c=\omega_s/\omega_c$.
This phenomenon also reveals the fact that the
transition from thermalization to localization can occur not only in many-body systems \cite{NandkishoreTL}, but
also dynamically in a single particle quantum system in contact with a reservoir \cite{Lin2016}.

\begin{figure}[b]
\includegraphics[width=12.0cm]{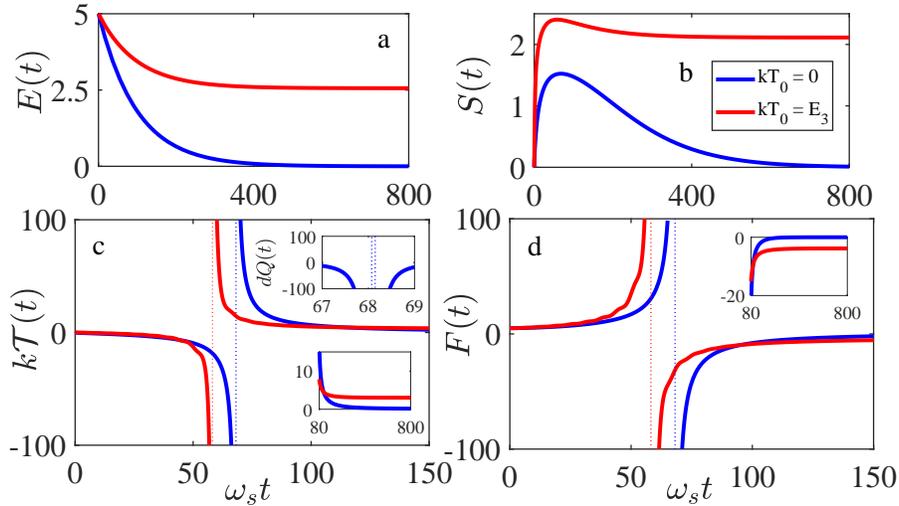}
\caption{\label{NT} {\bf Inflationary dynamics and dynamical quantum phase transition}.
The dynamics of (a) energy $E(t)$, (b) entropy $S(t)$, (c) dynamical temperature ${\cal T}(t)$,
and (d) free energy $F(t)$ are shown for the reservoir initially in vacuum state (blue curves) and
for the reservoir with initial thermal energy $kT_0=3 \hbar \omega_0$ (red curves). The system
is initially in the state $|E_{n_0}\rangle=|E_5\rangle$, and a weak system-reservoir coupling
strength $\eta=0.01\eta_c$ (blue) is taken.}
\end{figure}

\vskip 0.2cm
\noindent {\bf (iii).} {\it Dynamical quantum phase transition 
and quantum criticality with quantum-classical border}. Now we discuss a very interesting
situation occurring in the weak-coupling quantum regime. The quantum regime is defined
as the reservoir is initially in vacuum state ($T_0=0$) or the initial thermal energy of the reservoir is
less then the system initial energy: $kT_0 < E_0$, where the system is initially in the energy
eigenstate $|E_0\rangle$.
In this situation, the energy dissipation dominates the nonequilibrium process
so that the system energy always decreases in time (see Fig.~\ref{NT}a),
which is very different from the situation in Fig.~\ref{Weak}a
where the reservoir is in a relatively high temperature classical regime
$(kT_0 > E_0)$ and the  energy flows in an opposite way.
On the other hand, as shown in Fig.~\ref{NT}b the system entropy
(beginning with zero) will increase to a maximum value at an intermediate
time, then the entropy turns to decrease and eventually approaches to zero (the blue curve in Fig.~\ref{NT}b for $kT_0=0$ case).
This phenomenon can happen because the system is initially in a pure state (zero entropy),
then the system takes relaxation and goes eventually into the vacuum state (back to a pure state)
when all energy of the (small) system is eventually dissipated into the (large) reservoir. Thus, at the beginning and in the end of the
nonequilibrium process, the entropy of the system is zero. But during the nonequilibrium evolution,
the entropy must go up (increase) in time and reach the maximum at some intermediate time
(corresponding to the maximizing mixed state of the system), and then go down (decrease)
to zero again in the rest of relaxation process. On the other hand,
because the system energy always decreases, the nonequilibrium temperature which
starts from zero, must decrease to negative infinity as the entropy approaches to the maximum. It then
jumps from the negative infinite temperature to the positive infinite temperature at the maximum
entropy point, and then goes down as the entropy decreases, and eventually reaches to zero
temperature at the steady state, as shown by the blue curve in Fig.~\ref{NT}c.

This dramatic change of the dynamical temperature shows that for the nonequilibrium processes of open systems in the quantum regime,
the second law of thermodynamics cannot be always maintained (because of the appearance of a negative temperature and 
the decrease of the entropy).  But at the end, the entropy and the temperature both approach to zero,
as a manifestation of the third law of thermodynamics in equilibrium state.  Note that although the third
law of thermodynamics is hypothesized in classical thermodynamics, it cannot be demonstrated within
classical thermodynamics, namely the zero temperature is not reachable as the entropy cannot approach to zero
in the classical regime. This is because zero temperature locates in the very deep
quantum mechanical realm (corresponding to the situation of the system staying in the ground state).
Therefore as a criteria, a correct theory of quantum thermodynamics must capture
the generalized third law of thermodynamics, namely the entropy becomes zero as the temperature approach to
zero in quantum regime. Here we show explicitly how the third law of thermodynamics at equilibrium state
is obtained dynamically from quantum mechanics.

On the other hand,  the negative temperature in open quantum systems we find here, although it sounds to be
very strange, only appears in the
nonequilibrium region.  We find that this negative nonequilibrium temperature phenomenon always occurs when the system
is initially in a pure state with the condition that the system initial energy $E_0$ is large than the initial
thermal energy $kT_0$ of the reservoir. More importantly, we find that the occurrence of the negative
nonequilibrium temperature  is accompanied with a nontrivial behavior
of the entropy changing from increasing to decreasing, as shown in Fig.~\ref{NT}b-c. Here we would like to argue that
this unusual dynamics may provide a simple physical
picture to the origin of the universe inflation \cite{Guth1981}. Suppose that in the very beginning, our universe is a single
system staying in a pure quantum state \cite{Hawking} and nothing else, so that the reservoir is in vacuum.
Then the universe should evolve through a nonequilibrium process similar to that given in Fig.~\ref{NT}. The sudden jump
from the negative infinite temperature to the positive infinite temperature at the transition point
(corresponding to the entropy  varies from increase to decrease in time)
shown in Fig.~\ref{NT}c allows the system to suddenly generate an infinite amount of heat energy
$dQ = d[{\cal T}S] \rightarrow  \infty$ [i.e. an inflationary energy as a result of the maximizing information
(entropy) of the system]. At the same time, the free energy of the system is always increased
during the nonequilibrium evolution, but only at the transition point it is suddenly dropped. This sudden change
corresponds to deliver an inflationary work $dW\rightarrow \infty$ (i.e., an inflationary decrease of free energy of
the system, $dF\rightarrow -\infty$) to the reservoir,  see Fig.~\ref{NT}d.
This may provide naturally a simple physical process for the origin of big bang and the universe inflation
\cite{Guth1981,Hawking} which is a long-standing problem that has not been solved so far in quantum cosmology.
The continuous decrease of the entropy in the rest of the nonequilibrium
evolution may explain how the universe evolves to flatness.  Interestingly, an inflationary dynamics in Bose-Einstein
condensates (BEC) crossing a quantum critical point is recently observed \cite{Chin2018}. How to make the
above inflationary dynamics into a more realistic cosmology model remains for further
investigation.

\begin{figure}[t]
\includegraphics[width=15.0cm]{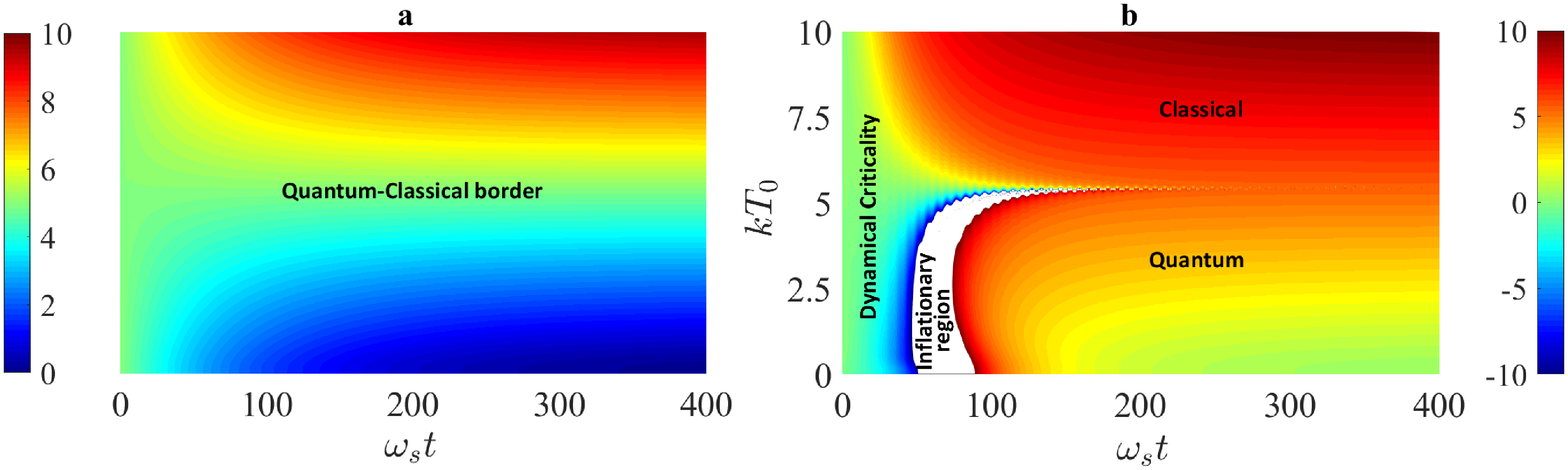}
\caption{\label{PhaseTransn} {\bf Dynamical quantum phase transition in quantum thermodynamics.}
Contour color-code plots of  (a) the dynamical energy $E(t)$ and (b) the
nonequilibrium temperature ${\cal T}(t)$ are presented as varying initial thermal energy $kT_0$ of
the reservoir. The other parameters are taken as $\eta=0.01~\eta_c$, $\omega_c = 5 \omega_s$, and
the system is considered to be in an initial state $|E_{n_0}\rangle=|E_5\rangle$. A dynamical quantum
phase transition occurs when initial thermal energy $kT_0$ of the reservoir is very close to the initial
system energy $E_{n_0}$.}
\end{figure}

Now if we change the initial state of the reservoir from vacuum to a thermal state with a low thermal energy,
$kT_0 < E_0$, the time-dependent entropy and the temperature of the system show the same behavior,
except for the steady state of the system which becomes now a thermal state with temperature $T_0$,
equilibrating to the reservoir, as shown by the red curves in Fig.~\ref{NT}.
Thus, when we change the initial thermal energy ($kT_0$)  of the reservoir from a lower value to a
higher value crossing the system initial energy $E_0$
(or alternatively change the system initial energy from $E_0 < kT_0$ to $E_0>
kT_0$ which is experimentally more feasible), we observe (see Fig.~\ref{PhaseTransn})
a nontrivial dynamical quantum phase transition when the thermal energy $kT_0$ of the reservoir is
very close to the initial system energy $E_0$. This quantum criticality provides indeed a
dynamical border separating the classical thermodynamics
from quantum thermodynamics, namely classical thermodynamics is only valid for a rather high
temperature regime ($kT_0 > E_0$), while in the quantum regime: $kT_0 < E_0$, only quantum thermodynamics
can capture thermodynamics phenomenon, manifested by the occurrence of
negative temperature and the decrease of entropy in the nonequilibrium process,
and it is finally also justified by the third law of thermodynamics in the steady-state limit.

\begin{figure}[t]
\includegraphics[width=14.0cm]{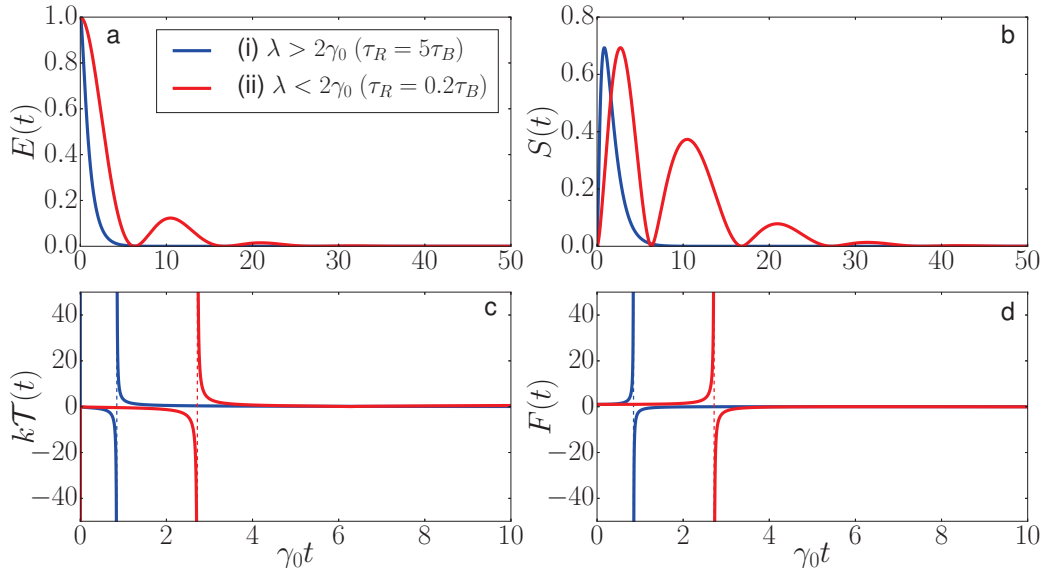}
\caption{\label{spinthermo} {\bf Nonequilibrium quantum thermodynamics of a two-level atomic system}.
We show the dynamics of nonequilibrium thermodynamic quantities at two physically distinct
regime (i) $\lambda > 2 \gamma_0$ and (ii) $\lambda < 2 \gamma_0$.
We plot (a) average energy $E(t)$ (b) entropy $S(t)$ (c) nonequiblibrium temperature ${\cal T}(t)$ and (d) free energy
$F(t)$, where the blue curves represent weak coupling regime (i) with $\tau_R=5\tau_B$, and the red curves indicate
the strong coupling regime (ii) with $\tau_R=0.2\tau_B$. The system is initially taken in the excited state, and the reservoir
is initially in vacuum state.}
\end{figure}

\vskip 0.8cm
\noindent
{\bf 3.~Application to a two-level atomic system.} To demonstrate the universality of the above finding in the
simple cavity system, we investigate next the real-time thermodynamics
of another widely studied open quantum system, namely a two-level atomic system (or a general spin-1/2 particle)
interacting with a bosonic reservoir, to see how thermodynamics emerges from quantum dynamics of spontaneous
decay process. The total Hamiltonian of the system plus reservoir is given by
\begin{eqnarray}
H = \hbar \omega_s \sigma_{+} \sigma_{-} + \sum_k \hbar \omega_k b_k^{\dagger} b_k +
\sum_k \hbar \left( g_k \sigma_{+} b_k + g_k^{\ast} \sigma_{-} b_k^{\dagger} \right),
\label{Spin}
\end{eqnarray}
where $\sigma_{+} = |1\rangle \langle 0|$ and $\sigma_{-} = |0 \rangle \langle 1|$ are the raising and
lowering operators of the Pauli matrix with ground state $|0\rangle$, excited state $|1\rangle$, and
transition frequency $\omega_s$. The reservoir Hamiltonian is described by a collection of infinite
bosonic modes with the bosonic operators $b_k^{\dagger}$ and $b_k$, and $g_k$ is the coupling strength between
the system and $k$th mode of the reservoir. We consider that the system is prepared initially in the excited
state $|1\rangle$ and the reservoir in a vacuum state. The time evolution of the reduced density matrix of the
two-level system is given by the exact master equation \cite{Garraway1997,SpinMaster}
\begin{eqnarray}
\label{JCmaster}
\frac{d}{dt} \rho(t) = - i \omega'_s(t,t_0) \left[ \sigma_{+} \sigma_{-} , \rho(t) \right]
+ \gamma(t,t_0) \left[ 2 \sigma_{-} \rho(t) \sigma_{+} - \sigma_{+} \sigma_{-} \rho(t) -
\rho(t) \sigma_{+} \sigma_{-} \right],
\end{eqnarray}
where the renormalized transition frequency $\omega'_s(t,t_0)= - \textrm{Im} \left({\dot u(t,t_0)}/u(t,t_0) \right)$
and the dissipation coefficient $\gamma(t,t_0) = - \textrm{Re} \left({\dot u(t,t_0)}/u(t,t_0) \right)$, which is the same as
that given in the general exact master
equation of Eq.~(\ref{Exact-ME}).  The Green function $u(t,t_0)$ satisfies also the same integro-differential equation as Eq.~(\ref{ide})
\begin{eqnarray}
\frac{d}{dt} u(t,t_0) + i\omega_s u(t,t_0)  + \int_{t_0}^{t} d\tau f(t,\tau) u(\tau,t_0) =0,
\label{ProbAmp}
\end{eqnarray}
where the two-time reservoir correlation function $f(t,\tau)$ is related to the spectral density
$J(\omega)$ of the reservoir as $f(t,\tau) \!= \!\!\int \!\! d\omega J(\omega) e^{-i\omega(t-\tau)} $.
Here we consider a Lorentzian spectral density,
\begin{eqnarray}
J(\omega) =  \frac{1}{2\pi}\frac{\gamma_0 \lambda^2}{(\omega_0 - \omega)^2 + \lambda^2},
\end{eqnarray}
for which the Green function $u(t,t_0)$ of Eq.~(\ref{ProbAmp}) can easily be solved with the exact solution
\begin{eqnarray}
u(t,t_0=0) = e^{(-i\omega_s -\frac{\lambda }{2})(t-t_0)} \left( \cosh \frac{\Gamma (t-t_0)}{2}
+ \frac{\lambda}{\Gamma} \sinh \frac{\Gamma (t-t_0)}{2}  \right),
\label{amptd}
\end{eqnarray}
where $\Gamma=\sqrt{\lambda^2 - 2 \gamma_0 \lambda}$, which has no localized bound state because
the spectral density covers the whole range of the frequency domain. The spectral width of the reservoir $\lambda$
is connected to the reservoir correlation time $\tau_B$ by the relation $\tau_B \approx \lambda^{-1}$. The
relaxation time scale $\tau_R$ over which the state of the system changes is related to the coupling
strength $\gamma_0$ by the relation $\tau_R \approx \gamma_{0}^{-1}$. The time evolved density
matrix $\rho(t)$ depicts the dissipative relaxation process of the system at far-from-equilibrium.
Once we have the density matrix $\rho(t)$, the time-dependent nonequilibrium thermodynamic quantities
are determined by the relations (\ref{vontropy}-\ref{work}). In Fig. \ref{spinthermo}, we show the
dynamics of nonequilibrium thermodynamic quantities, namely, internal energy $E(t)$, entropy $S(t)$,
temperature ${\cal T}(t)$ and the free energy $F(t)$. We consider two different regime of the system-environment
parameters: (i) $\lambda > 2 \gamma_0$, for which the reservoir correlation time is small compared to the
relaxation time ($\tau_B < \tau_R$) of the system and the behavior of $u(t,t_0)$ shows a Markovian
exponential decay (ii) $\lambda < 2 \gamma_0$, in this regime the reservoir correlation time $\tau_B$ is
large or comparable to the relaxation time scale $\tau_R$ where non-Markovian effects come into play
\cite{Bellemo07}.
We investigate the real-time thermodynamics of the system
in the above two physically distinct regimes (i) and (ii). In the weak coupling case with
$\tau_R=5\tau_B$, the energy $E(t)$ shows a monotonous decay due to dissipation,
as shown in Fig.~\ref{spinthermo}a (blue curve). On the other hand, Figure~\ref{spinthermo}b (blue curve) shows
that the entropy $S(t)$ increases due to the contact of the system to the reservoir,
and attains a maximum value at an intermediate time, henceforth the entropy starts decreasing and eventually
approaches to zero. This happens because the system is initially in a pure state (zero entropy), then the entropy
increases during the nonequilibrium evolution and reach the maximum corresponding to maximum mixing state
of the system, finally the entropy goes to zero again when the system relaxes into the vacuum state (pure state)
in order to equilibrate with the reservoir. The decrease in energy with increasing entropy results in a negative
nonequilibrium temperature of the system, as shown in Fig.~\ref{spinthermo}c. The nonequilibrium temperature ${\cal T}(t)$
decreased to negative infinity as the entropy approaches to the maximum, it then jumps from negative
infinite temperature to positive infinite temperature at the maximum entropy point, eventually the
system reaches to zero temperature at the steady state as shown in Fig.~\ref{spinthermo}c, the exactly
same as what we find in the cavity system in the weak-coupling quantum regime.

The physical origin of negative nonequilibrium temperature for this two-level atomic system is
also the same as that in cavity system: 
The free energy increases instead of decreasing
(see Fig.~\ref{spinthermo}d), an amount of work $dW$ is done on the system by the reservoir
resulting to an amount of heat $dQ$ dissipated from the system into the reservoir.
Hence, for this two-level system, the second law of thermodynamics also {\it cannot} be maintained 
when the system evolves from a pure state to a mixed state and then back to a pure state in the whole
spontaneous decay processes. Nevertheless, the entropy and the temperature both
dynamically approach to zero in the steady-state limit, as a manifestation of the third law of thermodynamics 
in equilibrium state in the quantum regime. In the strong coupling case with $\tau_R = 0.2 \tau_B$ (see the red 
curves of Fig.~\ref{spinthermo}), we show the 
nonequilibrium thermodynamic dynamics in the regime $\lambda < 2 \gamma_0$. In this regime,
the average energy $E(t)$ initially decreases, then it shows a small oscillation, demonstrating a
backflow of energy (red curve of Fig.~\ref{spinthermo}a) from the system into the reservoir, as a 
non-Markovian memory effect. We
see a similar dynamical behavior in the entropy $S(t)$ under this strong coupling, namely, the entropy
increases in the short time, then decreases and oscillates in later time (red curve of Fig.~\ref{spinthermo}b).
The nonequilibrium temperature and free energy in the strong coupling regime  (red curves of Fig.~\ref{spinthermo}c-d)
also show similar behaviors except a small shift in the time-scale.  The dynamical transition associated with the
inflationary dynamics is the same as we found in the single cavity system.

The concept of negative temperatures
has been discussed in the literature \cite{RamseyPR56,KleinPR56} which demand the system to be
in the equilibrium and also to have a finite
spectrum of energy states, bounded both from above and below. Such negative equilibrium temperature
has been realized in localized spin systems
\cite{PurcellPR51,OjaRMP97,KetterlePRL11}, where the discrete finite spectrum naturally provides
both lower and upper energy bounds. A similar situation is also found recently where a cloud of potassium
atoms is tuned to negative temperatures via quantum phase transition \cite{BraunScience}.
However, it is important to mention here that  the
physical origin or condition under which the negative temperature phenomenon occurs here in our cases is
quite distinct from the previous studies
\cite{RamseyPR56,KleinPR56,PurcellPR51,OjaRMP97,KetterlePRL11,BraunScience}. In previous studies
\cite{RamseyPR56,KleinPR56,PurcellPR51,OjaRMP97,KetterlePRL11,BraunScience}, it was
emphasized that in thermal equilibrium, the probability for a particle to occupy a state with energy
$E_n$ is proportional to the Boltzmann factor $\exp (- E_n/kT)$. For negative temperatures, the
Boltzmann factor increases exponentially with increasing $E_n$ and the high-energy states are then occupied
more than the low-energy states. Contrary to that, the emergence of negative temperature here we see
is dynamical, and it has the clear physical origin in the nonequilibrium dynamics evolution that the entropy
increases and then decreases as the system starts from a pure state, evolves into a maximum mixed state and then
returns back to a pure state (or a low mixed state) dominated by the spontaneous emissions
when the reservoir is initially in vacuum state (or an initial state with a very low thermal
energy).  Thus, our finding is indeed universal for other open quantum systems.

\vskip 0.5cm
\noindent
{\large \bf Conclusions and outlook}
\vskip 0.2cm
\noindent
Classical thermodynamics is built with the concept  of equilibrium states, where dynamics is absent and all thermodynamic
quantities are time-independent.   In this paper, we developed an exact nonequilibrium theory for quantum thermodynamics
validating for arbitrary quantum systems in contact with reservoirs at arbitrary time.  We generalize the
concept of temperature to nonequilibrium that depends on the details of quantum states of the system and
their dynamical evolution. We show that the exact nonequilibrium theory unravels the intimate connection between laws of thermodynamics
and their quantum origin. We explicitly and exactly solved the dynamical evolution of the quantum system interacting with a heat
reservoir. We demonstrated the emerging thermodynamics of the system through real-time dynamics of various
thermo-dynamic quantities, namely the time-dependent energy, entropy, temperature, free energy, heat and work in the weak
system-reservoir coupling regime under the situation much before the system approaches to thermal equilibrium,
and showed how the system dynamically approaches to a thermal equilibrium state asymptotically in the steady-state limit.
In the strong system-reservoir coupling regime, a revival dynamics in the nonequilibrium thermodynamic quantities is
observed, indicating back-reactions between the system and the environment as a non-Markovian memory effect, so that
equilibrium thermodynamics is unreachable, as noticed early by Anderson \cite{FanoAnd1} and justified recently
by one of us \cite{Complexity}.

Our most remarkable findings in this paper appear in the nonequilibrium thermodynamics with the reservoir thermal
energy $kT_0$ being less than the system initial energy, which is defined as the quantum regime.
Under this situation, we observe a dramatic change in the
thermodynamic behavior of the system that the system dissipates its energy into the reservoir in time, while the
entropy of the system first increases with the decrease of system energy so that the system is characterized with
a negative nonequilibrium temperature. After the
entropy reaches a maximum value at an intermediate time (corresponding to the maximized mixed state of the system),
the entropy decreases in the rest of nonequilibrium evolution.
Correspondingly, the nonequilibrium temperature jumps from the negative infinite to the positive infinite values at the maximum
entropy point, then decreases to the value equilibrating to the reservoir. This dramatic change of thermodynamic
behavior induces an inflationary heat energy in the system that suddenly delivers into the reservoir as an inflationary work
done by the system, which may provide a simple picture for the origin of big bang and universe inflation. The ending steady state manifests the
third law of thermodynamics in the deep quantum realm. We also observed a nontrivial dynamical quantum phase
transition when the thermal energy of the reservoir is very close to the system initial energy. The corresponding dynamical
criticality provides clearly a border separating quantum and classical thermodynamics. This dynamical quantum phase
transition is manifested through nonequilibrium quantum states that cannot be captured by equilibrium thermodynamics.

These new findings for quantum thermodynamics could be examined through BEC experiments or cavity QED experiments.
More specifically, one can experimentally explore the photon dynamics or boson dynamics via quantum non-demolition
measurement \cite{Exp1,Exp2} to justify these findings. For example, by preparing the cavity in a Fock state,
and sending sequences of circular Rydberg atoms through the photonic cavity in different environment, which
carry the cavity state information without destroying the cavity photon state. The cavity photon state can be measured
using an experimental setup similar to that given in Ref.~\cite{Exp1,Exp3}. Another way to observe the nontrivial
quantum thermodynamics we obtained is to measure the energy loss and its state evolution of a two-level atom
in cavity system. It may be also possible to use the BEC
systems \cite{Chin2018} to find these new features. On the other hand, it is also straightforward to
apply this quantum thermodynamics theory to electronic systems in nanostructures where the exact master equation
with quantum transport has been developed \cite{general,Yang2017}. Furthermore, our nonequilibrium theory for
quantum thermodynamics can provide a platform to probe the universal validity of second law of thermodynamics in
many other systems \cite{SecondLaw1,SecondLaw2,SecondLaw3,SecondLaw4}. We will also investigate various
fluctuation relations \cite{StrongNoneq3,FluctuationTheorem1,FluctuationTheorem2,FluctuationTheorem3} in the
quantum regime within this framework.

\vskip 0.5cm
\noindent
\textbf{Acknowledgments}

\noindent
WMZ acknowledges the support from the Ministry of Science and Technology of Taiwan under
Contract No.~NSC-102-2112-M-006-016-MY3.
MMA acknowledges the support from the Ministry of Science and Technology of Taiwan and
the Physics Division of National Center for Theoretical Sciences, Taiwan.

\ \\
\noindent \textbf{Author Contributions}

\noindent
MMA performed the detailed calculations and WMH performed the calculation for the two-level systems.~WMZ proposed the ideas and 
interpreted the physics.~MMA and WMZ both have contributed on discussions, physical formulation and writing the manuscript.

\ \\
\noindent\textbf{Competing financial interests}

\noindent The authors declare no competing financial interests.

\ \\
\noindent\textbf{Additional Information}

\noindent Correspondence and requests
for materials should be addressed to M. M. Ali (maniquantum@gmail.com) and W. M. Zhang (wzhang@mail.ncku.edu.tw).

\end{document}